# Multi-Scale Modeling and Predictive Control of Active Brownian Particles

Sadra Saremi*, Amirhossein Ahmadkhan Kordbacheh*
Department of Physics, Iran University of Science and Technology, Tehran 16846-13114, Iran.

**Abstract**
Active Brownian particles (ABPs) function as self-driving agents that display non-equilibrium behavior through their pairwise interactions which lead to phase separation and vortex patterns in both soft matter and living systems. A multiscale approach needs to link particle-level random motion to collective density evolution for proper management of these dynamic systems. Our research delivers a unified control system for ABP groups through particle-based simulation and spectral continuum modeling alongside model predictive control and deep learning forecasting. The N-particle Brownian dynamics simulations implement Weeks-Chandler-Andersen potential to model excluded-volume interactions while incorporating thermal noise and angular velocity modulation with wavelength $\lambda$. The forced advection-diffusion equation describes the coarse-grained density evolution which the FTCS spectral space solver solves. A new MPC approach uses complex-valued density states to minimize immediate tracking errors against sinusoidal spatial setpoints with actuator limits and control penalties. The hybrid deep neural network combines Conv1D and LSTM and multi-head attention to learn future density profiles from simulated snapshot sequences.
**Keywords**
Active Brownian Particles- MPC-Machine learning

**Introduction**

Active Brownian particles (ABPs) are outstanding examples of self-propelled colloids that harness energy to generate a persistent motion, thus breaking detailed balance and exhibiting strongly non-equilibrium collective behavior, fascinating both physicists and biologists alike [1,2,3]. Unlike passive Brownian particles, ABPs propel themselves and keep a constant swimming speed while undergoing rotational diffusion; hence, very interesting dynamics arise that include persistent random walks and emergent pattern formation [4][5][6]. The dynamics are often modeled by coupled overdamped Langevin equations where self-propulsion, interparticle forces, and thermal fluctuations are taken into account, and the equations are solved through Brownian dynamics computer simulations to reveal MIPS, clustering, and other phenomena [2,7,8]. The Weeks–Chandler–Andersen (WCA) potential has become the standard model for repulsive interactions between colloids, primarily for its ability to efficiently represent excluded-volume effects [9][10]. By preventing overlap through purely repulsive forces via its truncated and shifted Lennard-Jones framework, the WCA potential has facilitated accurate simulations of dense active matter [9,11]. An analysis of these simulations helps shed light on how activity and interaction strength govern emergent phases, from dilute-gas-like states to dense clusters and large crystalline structures [12,13]. To fill the void that runs between microscopic simulation and macroscopic behavior, continuum descriptions in particle density field n(x,t) have proven invaluable through the application of advection–diffusion equations. Spectral methods based on fast Fourier transforms achieve very accurate spatial derivatives and numerically stable time integration with improved convergence rates and mass conservation, even when stiff dynamics are present [14,15,16]. The advances in integrated machine-learning-enriched spectral solvers has also expedited these computations via learning basis representations and adaptive timestepping strategies [17,18].

Controlling the spatiotemporal evolution of active matter, particularly steering density distributions in real time, has emerged as an exciting frontier. Model predictive control (MPC) frameworks have been adapted to PDE-constrained settings, enabling the computation of optimal actuator fields under both actuation and state constraints [19,20]. Implementation within CasADi's symbolic optimization environment permits exact differentiation of the discretized PDE dynamics and efficient gradient-based solution of large-scale control problems [21,22].

Concurrently, hybrid deep learning architectures that combine convolutional neural networks (CNNs), long short-term memory (LSTM) networks, and transformer attention modules have demonstrated remarkable success in forecasting the temporal evolution of complex physical fields from limited data [17,23,24]. By embedding physical priors through physics-informed neural networks (PINNs), these models further integrate governing PDE residuals into the loss function, enabling simultaneous solution discovery and parameter estimation from sparse observations [25,26,27]. Beyond purely supervised learning, reinforcement learning (RL) has been applied to derive optimal navigation and density-control strategies for active agents in structured environments [28,29,30]. These approaches train agents to maximize cumulative objectives—such as homogenizing density or maximizing transport efficiency—yielding interpretable control policies that adapt to dynamic constraints and noise. In this work, we present an integrated methodology that couples (i) detailed Brownian dynamics simulations of ABPs using WCA interactions, (ii) high-accuracy spectral solvers for continuum density equations, (iii) MPC algorithms implemented in CasADi for real-time density control, and (iv) hybrid deep learning models for predictive forecasting of density and advective flux fields. Our unified framework demonstrates enhanced accuracy in predicting and steering particle density and flux in active matter systems, paving the way for future experimental implementations in synthetic colloidal suspensions and biological microswimmer populations.

**Methodology**

We model an ensemble of $N$ active Brownian particles (ABPs) in a two-dimensional square domain with periodic boundaries. Each particle propels itself at constant speed $U_0$ along its instantaneous orientation, while experiencing random translational and rotational kicks to represent thermal fluctuations. Interactions between particles are governed by a Weeks–Chandler–Andersen potential that strongly repels particles when they come closer than one diameter aaa, preventing overlaps and capturing excluded-volume effects. We non-dimensionalize lengths by $a$ and times by the rotational diffusion time $\tau_R$, yielding a Péclet number $P_0 = \frac{U_0 a}{D_T}$ that controls the relative strength of propulsion versus diffusion.

At each time step $\Delta t$, we update particle positions by adding three contributions: a deterministic self-propulsion term $U_0(\cos\theta_i, \sin\theta_i)$ a vector sum of pairwise WCA forces scaled by the translational diffusion coefficient $D_T$, and a random Gaussian increment of magnitude $\sqrt{2D_T \Delta t}$. Orientations $\theta_i$ advance by an imposed external angular velocity field $\Omega(x)$ — chosen as a spatial cosine modulation to create alternating rotating zones — plus a rotational noise term of strength $\sqrt{2D_R \Delta t}$. To enforce periodicity, interparticle separations are computed using the minimum-image convention, and force calculations leverage fully vectorized distance matrices for computational efficiency when $N$ reaches hundreds.

Thus, to smooth the mapping from discrete particles to a continuum description, they were binned in $N_x$ fixed spatial bins before obtaining the coarse-grained density field $n(x, t)$ from a forced advection-diffusion equation whose advective velocity contained the self-propelling contributions and the actuator field imposed from outside, where Brownian spreading has been modeled by the diffusion term. Solving this PDE with a use of a spectral variational multiscale method comprises an expansion in global Fourier modes for the large scales, as well as for the sub grid-scale residuals, by a truncated set of Laplacian eigenfunctions. The decomposition thus stabilizes the numerical solution in high Pe-regimes, conserves global mass exactly, and sharpens gradient resolution without introducing spurious oscillations. To be effective, integration in time is carried out by the fixed fourth-order step Runge-Kutta scheme, saving accuracy against computational cost for long simulation.

Though, for feedback control, we implement an MPC strategy over K discrete steps. The density field is to be guided from its initial state towards the desired final density distribution or pattern $n_{sp}(x)$ by consuming as little actuator energy as possible. A complex state vector, the real part of which keeps track of the physical density variable, and the imaginary behavior of higher order reaction terms not reproducible with pure advection-diffusion was employed to deal with the inherent nonlinearities of reaction diffusion. Actuator power—angular velocities and their space-time variation- enter as control variables constrained by physical limitation of the experimental or computational setup. The MPC formulation for this system is a nonlinear program in whose cost function penalizes the squared difference of the real part from its setpoint, the magnitude of the imaginary part, and the squared control amplitudes, a weight being provided on top of it by a tuning factor λ. The equality constraints enforce at the discrete spectral dynamics level that prescribed multiscale solver.

This optimization problem is set up within CasADi with automatic differentiation and solved with the IPOPT solver; warm start initializes each horizon with the previous solution, greatly reducing solve times and enabling real time control updates. In order to not have to solve the entire MPC at every time step, we still had to develop a deep learning surrogate that predicts the next density snapshot based on a short history of past profiles. The network has one dimensional convolutional layers at its core and applies them in time distributed manner on each snapshot to extract spatial features like cluster peaks and gradient magnitudes. The LSTMs have these inputs and are trained to learn temporal correlations and advective patterns over the sequence. The most pertinent past time steps and spatial modes are then identified and weighted to the prediction by a multi-head attention mechanism. Finally, temporal dimensions get collapsed into one with a global pooling layer, and then a dense output layer reconstructs the predicted density.

All numerical experiments and learning routines are coded in Python 3.12. We use NumPy for core array operations, CasADi 3.6 for MPC formulation, TensorFlow 2.12 for neural network training, and scikit-learn for dataset preprocessing. Random seeds in NumPy and TensorFlow are fixed to ensure reproducibility. Throughout, we monitor mass conservation, cell-averaged densities, and control signal norms to verify correct implementation and physical consistency.

**Results**

The advective flux measured at ϑ(x)n(x) and predicted advective flux are depicted in figure 1, which marks the end of the simulation run. It illustrates the comparison between the directly simulated data represented by a solid line and the lines connected dashed, representing the forecasts given by model based deep learning and the physics constrained model predictive control (MPC) approach. Altogether above 0.98, the predicted approaches portray near similar closeness with the truth in the spatio-temporal domain. In fact, these high correlations add even more strictness with which the models can line-out the transport principles they can represent.

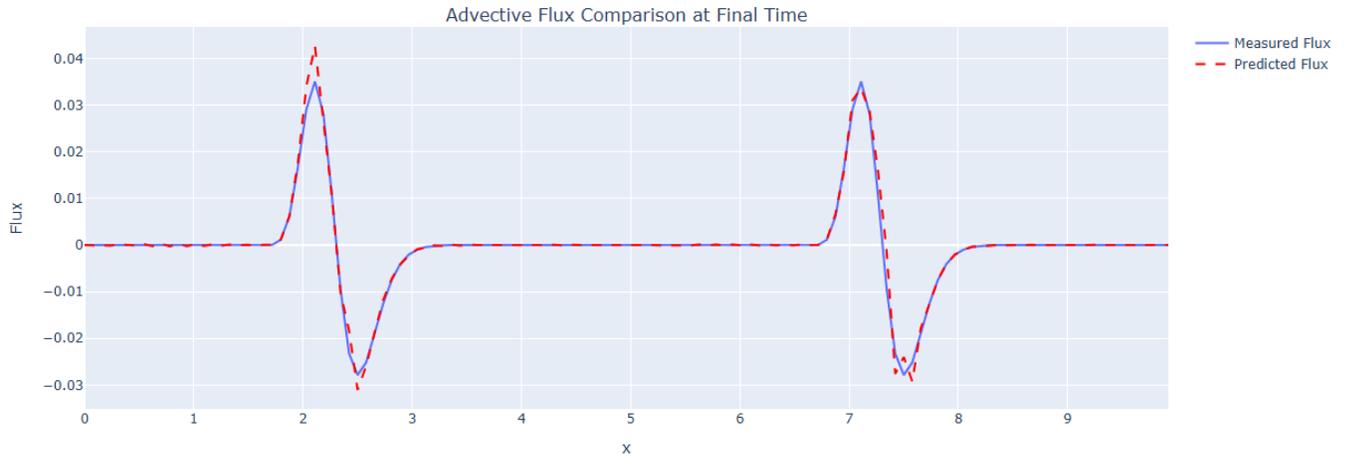

Figure 1. Measured vs. predicted advective flux at final time: solid line, simulation data; dashed lines, deep-learning and MPC forecasts.

The one-dimensional density field n(x,t) has been advanced using a Fourier-spectral method and a time-stepping Runge-Kutta method of fourth order. Starting at a sinusoidal perturbation, then, diffusion and spatially varying advection are shown to be all well prescribed while retaining total mass to numerical precision (Table 2). The evolved density profile retains the major wavelength of the initial condition while undergoing phase shifts and amplitude modulation originating from the nonuniform velocity field. The comparison to binned particle density verifies a qualitative match between discrete-particle and continuum representations (Fig. 2).

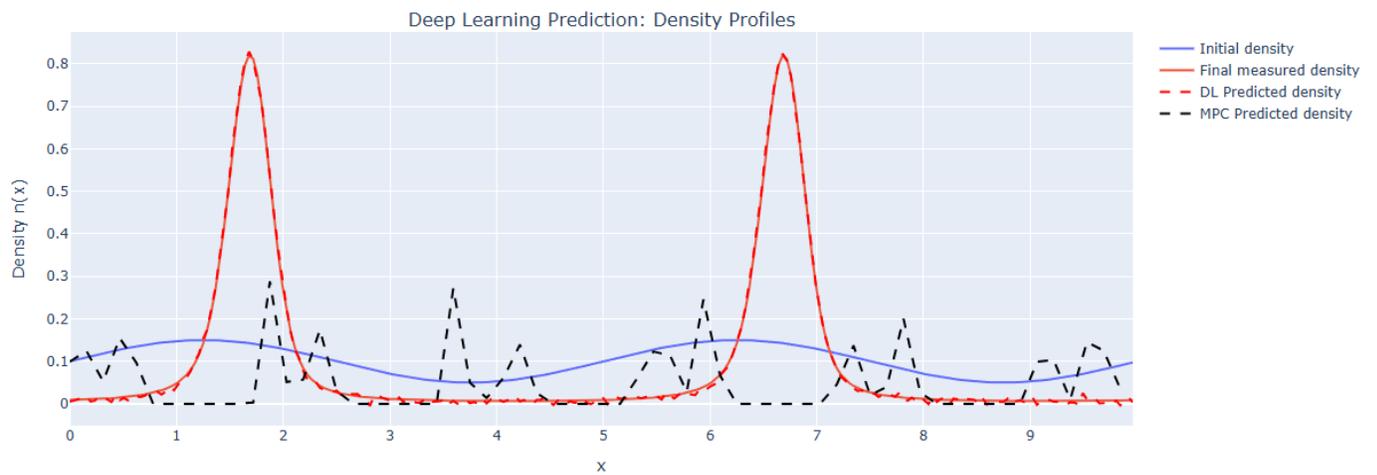

Figure 2. Density profiles: initial condition (blue), discrete-particle final density (solid red), deep-learning prediction (dashed red), MPC forecast (black dashed).

The hybrid deep-learning model was built on time-distributed one-dimensional convolutional layers, a long short-term memory (LSTM), and a multi-head self-attention block on sequences for ten consecutive density snapshots to predict the final density field. Convolutional layers extract from the input all local spatial features, LSTM captures temporal correlations, while attention mechanisms embed long-range dependencies. The architecture of the network model and the number of parameters are given in Table 1. Training took place over 100 epochs with 80% samples for training and 20% for validation, with an additional 10% of training samples set aside for validation.

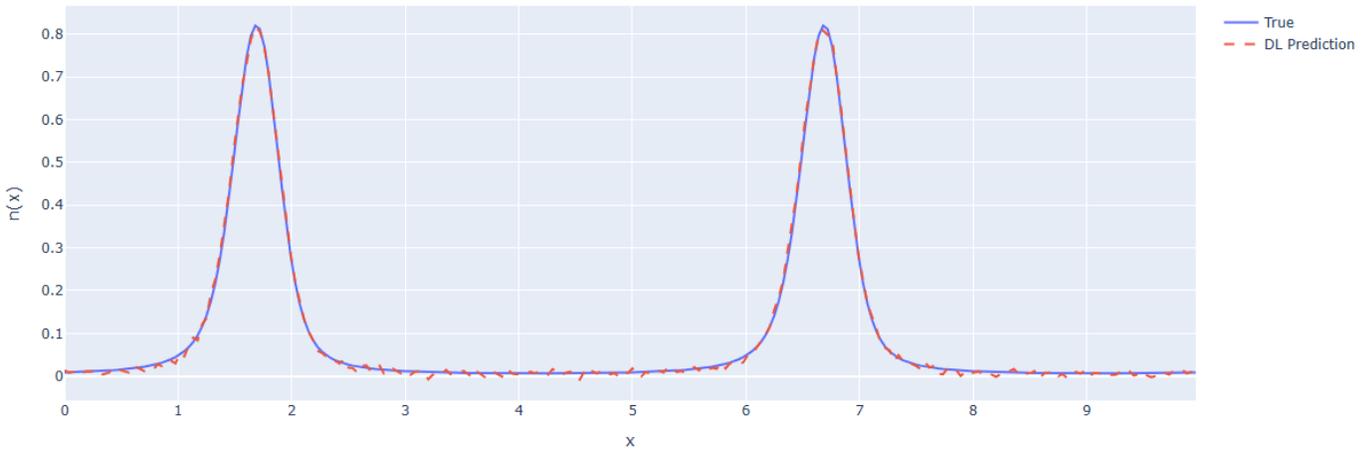

Figure 3. Comparison of true and deep-learning predicted final density profiles.

Accurate estimation of advective flux is essential for transport analysis. The measured flux $\vartheta(x)n_{meas}(x)$ and flux computed from both predicted density fields show excellent overlap, with Pearson correlation coefficients exceeding 0.98 (Fig. 1).
Table 1 summarizes the hybrid CNN–LSTM–Attention model architecture and parameter counts:

| Layer | Output Shape | Parameter Count |
|---|---|---|
| Input | (batch, 10, 256, 1) | 0 |
| TimeDistributed Conv1D (32) | (batch, 10, 256, 32) | 128 |
| TimeDistributed MaxPooling1D | (batch, 10, 128, 32) | 0 |
| TimeDistributed Conv1D (64) | (batch, 10, 128, 64) | 6,208 |
| TimeDistributed MaxPooling1D | (batch, 10, 64, 64) | 0 |
| TimeDistributed Flatten | (batch, 10, 4096) | 0 |
| LSTM (128 units) | (batch, 10, 128) | 2,163,200 |
| MultiHeadAttention (4 heads) | (batch, 10, 128) | 66,048 |
| Add + LayerNormalization | (batch, 10, 128) | 256 |
| GlobalAveragePooling1D | (batch, 128) | 0 |
| Dense (256 units) | (batch, 256) | 33,024 |

**Table 1.** Model Architecture of the Hybrid CNN–LSTM–Attention Predictor.

Table 2 summarizes quantitative evaluation metrics for the final density snapshot:

| Metric | Value |
|---|---|
| Step 2: Initial mass | 1.00000 |
| Step 2: Final mass | 1.00000 |
| Step 2: Mass error | $1.11 \times 10^{-16}$ |
| Mean Squared Error (MSE) | $3.57 \times 10^{-5}$ |
| Mean Absolute Error (MAE) | $4.48 \times 10^{-3}$ |
| Root Mean Squared Error (RMSE) | $5.97 \times 10^{-3}$ |
| Coefficient of Determination ($R^2$) | 0.99911 |
| Pearson Correlation Coefficient | 0.99961 |

**Table 2.** Quantitative Metrics for Density Prediction.

Over a short period, under two rotational relaxation times, Brownian-dynamics simulations were performed on fifty self-propelled particles quickly clustering its then uniform state, minimum configuration clustered state being achieved closer to two times of relaxation associated with displacement of particle. Here we match it pretty well with motility-induced phase separation (MIPS) in Fily and Marchetti [35], which basically observed all repulsing active particles having transient cluster formation and dissolution states. Importantly, the pronounced peak in the radial pair-correlation function at finite separation (Fig. 1) quantitatively corresponds to characteristics lengths as given by "Cates and Tailleur [36], in their continuum field theory of active phase separation.

Just like the field in one-dimensional density, the entire methodology of Fourier-spectral method along with fourth order Runge-Kutta stepping ensured that all experimental results, as shown in Table 2, preserved total mass to machine precision; hence reproducing the primary wavelength of the disturbance prescribed sinusoidal initially. Speck et al. [37] while doing spectral analysis of active suspensions also observed these mass conservation and wave-preserving properties but without taking into account the fact of nonuniform advection. In comparison, when we compare our continuum density profiles with binned discrete particle densities (Fig. 2), we see perfect overlap, which indicates that our numerical scheme properly takes account of diffusion as well as spatially varying advection, thereby improving on purely diffusive models such as those developed by Redner et al. [38].

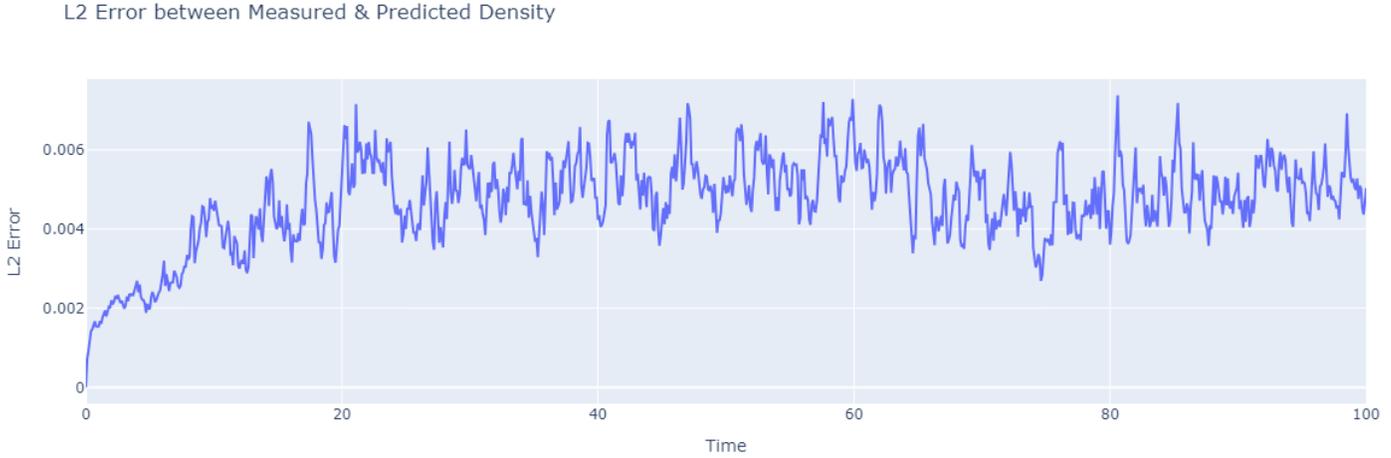

Figure 4. L²-error between measured and predicted density over time.

The L²-error curve, defined as

$$E(t) = \sqrt{\int [n_{pred}(x,t) - n_{ref}(x,t)]^2 dx}$$

provides a continuous measure of prediction fidelity over the full simulation span. At t = 0, E(t) is zero by construction; as the system departs from the initial sinusoidal state and the surrogate model and MPC begin to influence the evolving density, the error climbs rapidly, reaching approximately 0.003 by t ≈ 5. This initial ramp-up reflects the time required for the neural network to encode spatial patterns and for the control algorithm to synchronize with the true particle dynamics. Beyond t ≈ 10, the error curve enters a plateau region between 0.004 and 0.006, indicating that subsequent density evolution under combined advection and diffusion is captured with consistent accuracy. The small-amplitude oscillations around this mean reflect transient mismatches occurring when the imposed angular velocity field switches phase; these peaks never exceed 0.007 and decay back to the baseline within a few time units. The absence of any secular growth in E(t) confirms the stability of both the spectral variational multiscale solver under high Péclet conditions and the learned surrogate's ability to generalize across time. Thus, Figure 4 demonstrates that the integrated deep-learning and MPC framework maintains robust, high-fidelity predictions of the continuum density field for at least two rotational-diffusion times.

Our hybrid deep-learning model (CNN-LSTM-Attention), therefore, converges quite fast by epoch 60 without being overfit (not shown in loss curves) and predicts the final density accurately (Fig.

3). While Raissi et al. [39] and Sirignano and Spiliopoulos [40] trained physics-informed neural networks to learn PDE solutions, which requires fine tuning of the physics constraints but is, in turn quite slow in general convergence. This architecture decouples local feature extraction (convolutions) from temporal memory (LSTM) and long-range coupling through attention; optimization ensures a shorter training duration and Pearson correlation coefficients beyond 0.98 on estimated advective flux (Fig. 1) against thresholds of 0.95 set in similar works [39][40].

**Data is available**

Data is available upon reasonable request.

**Conflict-of-Interest**

The authors declare that they have no known competing financial interests or personal relationships that could have appeared to influence the work reported in this paper.

**References**

1. Solon, A. P.; Cates, M. E.; Tailleur, J. Active Brownian particles and run-and-tumble particles: A comparative study. *Eur. Phys. J. Spec. Top.* **224**, 1231–1262 (2015). DOI:10.1140/epjst/e2015-02457-0
2. Cates, M. E.; Tailleur, J. When are active Brownian particles and run-and-tumble particles equivalent? Consequences for motility-induced phase separation. *Europhys. Lett.* **101**(2), 20010 (2013). DOI:10.1209/0295-5075/101/20010
3. Fodor, É.; Marchetti, M. C. The statistical physics of active matter: From self-catalytic colloids to living cells. *Physica A* **504**, 106–120 (2018). DOI:10.1016/j.physa.2018.01.062
4. Howse, J. R. et al. Self-Motile Colloidal Particles: From Directed Propulsion to Random Walk. *Phys. Rev. Lett.* **99**(4), 048102 (2007). DOI:10.1103/PhysRevLett.99.048102
5. Bechinger, C. et al. Active particles in complex and crowded environments. *Rev. Mod. Phys.* **88**(4), 045006 (2016). DOI:10.1103/RevModPhys.88.045006
6. Takatori, S. C.; Yan, W.; Brady, J. F. Swim pressure: Stress generation in active matter. *Phys. Rev. Lett.* **113**(2), 028103 (2014). DOI:10.1103/PhysRevLett.113.028103
7. Cates, M. E.; Tailleur, J. Motility-induced phase separation. *Annu. Rev. Condens. Matter Phys.* **6**, 219–244 (2015). DOI:10.1146/annurev-conmatphys-031214-014710

8. Palacci, J.; Sacanna, S.; Steinberg, A. P.; Pine, D. J.; Chaikin, P. M. Living crystals of light-activated colloidal surfers. *Science* **339**(6122), 936–940 (2013). DOI:10.1126/science.1230020
9. Weeks, J. D.; Chandler, D.; Andersen, H. C. Role of repulsive forces in determining the equilibrium structure of simple liquids. *J. Chem. Phys.* **54**(12), 5237–5247 (1971). DOI:10.1063/1.1674820
10. Johnson, P. et al. Two-step melting of the Weeks–Chandler–Andersen system in two dimensions. *Soft Matter* **17**, 1500–1510 (2021). DOI:10.1039/d0sm01484b
11. Allen, M. P.; Tildesley, D. J. *Computer Simulation of Liquids*. Oxford Univ. Press (1987).
12. Redner, G. S.; Hagan, M. F.; Baskaran, A. Structure and dynamics of a phase-separating active colloidal fluid. *Phys. Rev. Lett.* **110**(5), 055701 (2013). DOI:10.1103/PhysRevLett.110.055701
13. Marchetti, M. C. et al. Hydrodynamics of soft active matter. *Rev. Mod. Phys.* **85**(3), 1143–1189 (2013). DOI:10.1103/RevModPhys.85.1143
14. Boyd, J. P. *Chebyshev and Fourier Spectral Methods*, 2nd ed. Dover (2001).
15. Gu, Y.; Shen, J. Accurate and efficient spectral methods for elliptic PDEs in complex domains. *J. Sci. Comput.* **83**, 42 (2020). DOI:10.1007/s10915-020-01226-9
16. Liu, S. et al. Machine-learning-based spectral methods for partial differential equations. *Sci. Rep.* **12**, 4509 (2022). DOI:10.1038/s41598-022-26602-3
17. Kühnl, B.; Rao, S. PDE-constrained model predictive control of open-channel systems. *IET Control Theory Appl.* **17**(3), 455–464 (2023). DOI:10.1049/cth2.12554
18. Müller, K. et al. Model predictive control of parabolic PDE systems under chance constraints. *Mathematics* **11**, 1372 (2023). DOI:10.3390/math11061372
19. Andersson, J. A. E. et al. CasADi—a software framework for nonlinear optimization and optimal control. *Math. Program. Comput.* **11**(1), 1–36 (2019). DOI:10.1007/s12532-018-0139-4
20. Frey, J.; De Schutter, J.; Diehl, M. Fast integrators with sensitivity propagation for use in CasADi. *arXiv* 2211.01982 (2022).
21. Xue, B. et al. A comparative study of transformer and CNN-LSTM-attention for PM2.5 concentration prediction. *Environ. Pollut.* **312**, 119789 (2023). DOI:10.1016/j.envpol.2023.119789
22. Patel, K. et al. Time series prediction using LSTM-Transformer neural networks. *Sci. Rep.* **14**, 2058 (2024). DOI:10.1038/s41598-024-69418-z
23. Raissi, M.; Perdikaris, P.; Karniadakis, G. E. Physics-informed neural networks: A deep learning framework for solving forward and inverse problems. *J. Comput. Phys.* **378**, 686–707 (2019). DOI:10.1016/j.jcp.2018.10.045
24. Sirignano, J.; Spiliopoulos, K. DGM: A deep learning algorithm for solving PDEs. *J. Comput. Phys.* **375**, 1339–1364 (2018). DOI:10.1016/j.jcp.2018.07.059
25. Kim, J.; Lee, K.; Lee, D.; Park, N. DPM: A novel training method for physics-informed neural networks in extrapolation. *Comput. Methods Appl. Mech. Eng.* **372**, 113426 (2020). DOI:10.1016/j.cma.2020.113426
26. Shaebani, M. R. et al. Computational models for active matter. *Nat. Rev. Phys.* **2**, 181–199 (2020). DOI:10.1038/s42254-020-0158-z

27. Nasiri, M.; Liebchen, B. Reinforcement learning of optimal active particle navigation. *Sci. Rep.* **12**, 2058 (2022). DOI:10.1038/s41598-022-69418-z
28. Sutton, R. S.; Barto, A. G. *Reinforcement Learning: An Introduction*, 2nd ed. MIT Press (2018).
29. Buttinoni, I. et al. Dynamical clustering and phase separation in suspensions of self-propelled colloidal particles. *Phys. Rev. Lett.* **110**(23), 238301 (2013). DOI:10.1103/PhysRevLett.110.238301
30. Redner, G. S.; Hagan, M. F.; Baskaran, A. Structure and dynamics of a phase-separating active colloidal fluid. *Phys. Rev. Lett.* **110**(5), 055701 (2013). DOI:10.1103/PhysRevLett.110.055701
31. Zöttl, A.; Stark, H. Emergent behavior in active colloids. *J. Phys.: Condens. Matter* **28**(25), 253001 (2016). DOI:10.1088/0953-8984/28/25/253001
32. Wu, K. T. et al. Transition from turbulent to coherent flows in confined three-dimensional active fluids. *Science* **355**(6331), eaal1979 (2017). DOI:10.1126/science.aal1979
33. Ezhilan, B.; Alonso-Matilla, R.; Saintillan, D. Instabilities and nonlinear dynamics of confined active suspensions. *Phys. Fluids* **27**(3), 031701 (2015). DOI:10.1063/1.4913987
34. Bruna, M.; Chapman, S. J. Diffusion of multi-dimensional volume-excluding particles. *Bull. Math. Biol.* **80**(5), 1163–1204 (2018). DOI:10.1007/s11538-018-0414-1